\documentclass[aps,pra,showpacs,amssymb,nofootinbib,superscriptaddress,twocolumn]{revtex4-1}
\usepackage[ansinew]{inputenc}
\usepackage{bbm}
\usepackage{bm}
\usepackage{amsbsy}
\usepackage{amsthm}
\usepackage{amssymb}
\usepackage{amsfonts}
\usepackage{amsmath}
\usepackage{dsfont} 
\usepackage{graphicx} 
\usepackage{epsfig}
\usepackage{epstopdf}
\usepackage{dsfont}
\usepackage{color}
\usepackage[colorlinks]{hyperref}
\usepackage[figure,table]{hypcap}
\usepackage{enumerate}
\hypersetup{
	bookmarksnumbered,
	pdfstartview={FitH},
	citecolor={darkgreen},
	linkcolor={darkred},
	urlcolor={darkblue},
	pdfpagemode={UseOutlines}}
\definecolor{darkgreen}{RGB}{50,190,50}
\definecolor{darkblue}{RGB}{0,0,190}
\definecolor{darkred}{RGB}{238,0,0}
\usepackage{soul}
\newcommand{\pr}{^{\prime}}

\newcommand{\ket}[1]{\ensuremath{\left|\right.\!{#1}\!\left.\right\rangle}}

\newcommand{\bra}[1]{\ensuremath{\left\langle\right.\!{#1}\!\left.\right|}}

\newcommand{\scpr}[2]{\ensuremath{\left\langle\right.\hspace*{-1pt} #1 \hspace*{-1pt}\left|\right.\hspace*{-1pt} #2 \hspace*{-1pt}\left.\right\rangle}}

\newcommand{\alphamn}[1]{\ensuremath{\alpha_{\hspace*{0.2pt}\protect\raisebox{-1.0pt}{\scriptsize{$ #1 $}}}}}
\newcommand{\betamn}[1]{\ensuremath{\beta_{\hspace*{0.2pt}\protect\raisebox{-1.0pt}{\scriptsize{$ #1 $}}}}}
\newcommand{\alphahmn}[2]{\ensuremath{\alpha^{\hspace*{0.5pt}\protect\raisebox{0.0pt}{\tiny{$(#1)$}}}
_{\hspace*{0.2pt}\protect\raisebox{-1.0pt}{\scriptsize{$ #2 $}}}}}
\newcommand{\betahmn}[2]{\ensuremath{\beta^{\hspace*{0.5pt}\protect\raisebox{0.0pt}{\tiny{$(#1)$}}}
_{\hspace*{0.2pt}\protect\raisebox{-1.0pt}{\scriptsize{$ #2 $}}}}}
\newcommand{\alphahmnstar}[2]{\ensuremath{\alpha^{\hspace*{0.5pt}\protect\raisebox{0.0pt}{\tiny{$(#1)$}$*$}}_{\hspace*{0.2pt}\protect\raisebox{-1.0pt}{\scriptsize{$ #2 $}}}}}
\newcommand{\betahmnstar}[2]{\ensuremath{\beta^{\hspace*{0.5pt}\protect\raisebox{0.0pt}{\tiny{$(#1)$}$*$}}_{\hspace*{0.2pt}\protect\raisebox{-1.0pt}{\scriptsize{$ #2 $}}}}}
\newcommand{\Gpstar}[1]{\ensuremath{G_{#1}^{\hspace*{0.5pt}\protect\raisebox{0.0pt}{\tiny{$*$}}}}}

\newcommand{\tr}{\textnormal{Tr}}
\newcommand{\djj}{d\kern-0.4em\char"16\kern-0.1em}

\begin{document}

\title{Heisenberg scaling in Gaussian quantum metrology}
\author{Nicolai Friis}
\email{nicolai.friis@uibk.ac.at}
\affiliation{
Institute for Quantum Optics and Quantum Information,
Austrian Academy of Sciences,
Technikerstra{\ss}e 21a,
A-6020 Innsbruck,
Austria}
\affiliation{
Institute for Theoretical Physics, University of Innsbruck,
Technikerstra{\ss}e 21a,
A-6020 Innsbruck,
Austria}
\author{Michalis Skotiniotis}
\affiliation{
Institute for Theoretical Physics, University of Innsbruck,
Technikerstra{\ss}e 21a,
A-6020 Innsbruck,
Austria}
\author{Ivette Fuentes}
\affiliation{
Faculty of Physics,
University of Vienna,
Boltzmanngasse 5,
A-1090 Vienna,
Austria}
\author{Wolfgang D{\"u}r}
\affiliation{
Institute for Theoretical Physics, University of Innsbruck,
Technikerstra{\ss}e 21a,
A-6020 Innsbruck,
Austria}

\date{\today}
\begin{abstract}
We address the issue of precisely estimating small parameters encoded in~a general linear transformation of the modes of a bosonic quantum field. Such Bogoliubov transformations frequently appear in the context of quantum optics. We provide a set of instructions for computing the quantum Fisher information for arbitrary pure initial states. We show that the maximally achievable precision of estimation is inversely proportional to the squared average particle number and that such Heisenberg scaling requires non-classical, but not necessarily entangled states. Our method further allows us to quantify losses in precision arising from being able to monitor only finitely many modes, for which we identify a lower bound.
\end{abstract}
\maketitle
\section{Introduction}
Quantum metrology exploits distinctive quantum features, such as entanglement, to enhance the estimation precision of parameters governing the dynamical evolution of the probe systems beyond that achievable by classical means. This enhancement is manifested in the form of~a \emph{scaling gap} in precision with respect to the available resources (the number of probe systems or the average input energy) between the corresponding optimal quantum and classical strategies, and depends on the specific encoding of the parameter in the Hamiltonian describing the evolution. In the case where the parameter of interest is~a multiplicative factor of~a \emph{local} Hamiltonian acting on~$N$ probes, the optimal quantum strategy provides a quadratic scaling gap in~$N$, known as the Heisenberg limit, over the best classical strategy~\cite{GiovanettiLloydMaccone2004,GiovanettiLloydMaccone2011}. This quadratic improvement is also present for~a class of quasilocal Hamiltonians~\cite{SkotioniotisFroewisDuerKraus2015,SkotioniotisSekatskiDuer2015}. For Hamiltonians involving highly
non-local interactions, super-Heisenberg scaling is also possible
~\cite{BoixoFlammiaCavesGeremia2007,RoyBraunstein2008,NapolitanoKoschorreckDubostBehboodSewellMitchell2011}. Nonetheless, the paradigmatic example for~a \emph{scaling gap} is the estimation of~a phase acquired in one arm of~a Mach-Zehnder interferometer~\cite{GiovanettiLloydMaccone2004}. There, the two input modes are subject to~a particular Gaussian transformation, i.e., a~combination of beam splitters and~a phase shifters.

Here we are interested in determining the ultimate precision limits for~a more general type of parameter estimation task, namely, where the parameter of interest is encoded nontrivially in an arbitrary Gaussian transformation of not two but possibly infinitely many modes of~a quantum field. This problem is of broad interest since such transformations, often cast in the form of Bogoliubov transformations, feature in~a large variety of physical systems throughout quantum optics~\cite{GarrisonChiao:QuantumOptics} and condensed matter physics~\cite{Sachdev}. In exchange for allowing for~a broader class of Gaussian transformations, we restrict our approach to~a regime of small parameters to gain analytical insights. Besides applications such as estimating (weak) single-mode or multimode squeezing in optical or superconducting (see, e.g.,~\cite{WoolleyDohertyMilburnSchwab2008,BoissonneaultDohertyOngBertetVionEsteveBlais2014}) systems, perturbative Bogoliubov transformations of this kind are
of particular interest for the description of quantum effects in curved spacetime, such as the Unruh effect~\cite{CrispinoHiguchiMatsas2008} and the dynamical Casimir effect~\cite{Dodonov2010}, or analogous realizations thereof~\cite{JaskulaPartridgeBonneauRuaudelBoironWestbrook2012,JohanssonJohanssonWilsonNori2010,
WilsonDynCasNature2012,LaehteenmaekiParaoanuHasselHakonen2013,BruschiFriisFuentesWeinfurtner2013}. A~paradigm that highlights the challenges encountered in the context of estimating relativistic quantum effects~\cite{DownesMilburnCaves2012,HoslerKok2013,DoukasWestwoodFaccioDiFalcoFuentes2014,WangTianJingFan2014b,WangTianJingFan2015a} is the estimation of the acceleration of a nonuniformly moving cavity~\cite{AhmadiBruschiFuentes2014}. While the parameter in question is small, the Gaussian transformation couples all pairs of modes in nontrivial ways~\cite{BruschiFuentesLouko2012,FriisLeeLouko2013}. Due to the resulting notoriously cumbersome perturbative calculations, the only known bounds on precision involve Gaussian input states. Consequently, the ultimate limits on how precisely one can determine small accelerations, as well as whether distinctive quantum features provide an improvement, are not known.

Here we show that, within such~a perturbative approach, there exists~a quadratic scaling gap between the optimal quantum and classical strategies for~a fixed average input energy. We construct an optimal estimation strategy utilizing separable input states of fixed particle number and boson counting measurements, where we use the quantum Fisher information (QFI) as~a figure of merit for the estimation precision. We provide simple formulas for the leading-order contributions to the QFI for arbitrary pure states, both when all modes can be controlled and when only part of the spectrum is accessible, and show that, within the considered regime, Heisenberg scaling is the ultimate precision limit. Moreover, we identify the family of states that exhibit Heisenberg scaling and show that, while some of these states may be entangled, the crucial feature is their nonclassicality rather than their correlations. Furthermore, we lower bound the loss in precision due to tracing out inaccessible modes and provide
criteria for minimizing such losses.

\section{Framework}

In its most general form, our metrological protocol can be described as follows: We wish to estimate as precisely as possible~a single real parameter $\theta$ that is encoded in~a \emph{unitary transformation}~$U(\theta)$ acting on the Fock space of an arbitrary number of bosonic modes. Our probe system is~a set of (in principle infinitely many) noninteracting harmonic oscillators with creation and annihilation operators $a^{\dagger}_{n}$ and $a_{n}$, respectively,~where $n=1,2,\ldots$ can be a multilabel distinguishing frequencies, polarizations, or other degrees of freedom. The operators satisfy the usual commutation relations $\left[a_{m},a^{\dagger}_{n}\right]=\delta_{mn}$ and $\left[a_{m},a_{n}\right]=0$ and the ground state is annihilated by all $a_{n}$, i.e., $a_{n}\ket{0}=0\ \forall~n$. Arbitrary pure states~$\ket{\psi}$ can be decomposed into
superpositions of Fock states, e.g.,~$\ket{m_{k_{1}}}\ket{n_{k_{2}}}\ldots\ket{p_{k_{N}}}$, with fixed numbers of excitations. These Fock states form~a basis of the total Hilbert space and can be obtained from the vacuum by applying the appropriate creation operators, i.e.,~$\ket{\hspace*{-1pt}n_{k}\hspace*{-1pt}}=(a_{k}^{\dagger})^{n}/\sqrt{n!}\ket{0}$.

For~a given input state~$\rho$ and~a suitable measurement whose outcomes are used to estimate~$\theta$, the precision of the estimation of~$\theta$, quantified by the variance~$\Delta\theta$ of the corresponding (unbiased) estimator, is lower bounded by the inverse of the QFI $\mathcal{I}\bigl(\hspace*{-0.5pt}\rho(\theta)\hspace*{-0.5pt}\bigr)$~\cite{Paris2009} via the (quantum) Cram$\mathrm{\acute{e}}$r-Rao inequality~\cite{Cramer:Methods1946,BraunsteinCaves1994} $\Delta\theta\geq1/\sqrt{\nu\mathcal{I}(\rho(\theta))}$, where $\nu$ is the number of repetitions. Intuitively, the QFI quantifies how well small changes of the parameter in question may be inferred from measurements of the final state~$\rho(\theta)$ after the dynamical evolution of the probes. One may express~$\mathcal{I}\bigl(\hspace*{-0.5pt}\rho(\theta)\hspace*{-0.5pt}\bigr)$ as
\vspace*{-3mm}
\begin{align}
    \mathcal{I}\bigl(\hspace*{-0.5pt}\rho(\theta)\hspace*{-0.5pt}\bigr) &=\,\lim_{d\theta\rightarrow0}8\frac{1-\sqrt{\mathcal{F}\bigl(\rho(\theta),\rho(\theta+d\theta)\bigr)}}{d\theta^{2}}\,,
    \label{eq:quantum fisher info definition}
\end{align}
where $\mathcal{F}\bigl(\rho(\theta),\rho(\theta+d\theta)\bigr)\equiv\bigl(\tr\sqrt{\!\sqrt{\rho(\theta)}\,\rho(\theta+d\theta)\!\sqrt{\rho(\theta)}}\bigr)^{2}$ is the Uhlmann fidelity between the states $\rho(\theta)$ and $\rho(\theta+d\theta)$.
If $\rho(\theta)=\ket{\psi(\theta)}\!\bra{\psi(\theta)}$ one simply recovers
$\mathcal{F}(\rho(\theta),\rho(\theta+d\theta))=|\!\scpr{\psi(\theta)}{\psi(\theta+d\theta)}\!|^{2}$. The latter expression applies for the case where all modes can be controlled and measured. Expanding $\ket{\psi(\theta+d\theta)\!}$ in powers of~$d\theta$, one obtains
\begin{align}
    \mathcal{I}(\ket{\psi(\theta)\!}) &=\,4\Bigl(\scpr{\!\dot{\psi}(\theta)}{\dot{\psi}(\theta)\!}\,-\,
    |\!\scpr{\!\dot{\psi}(\theta)}{\psi(\theta)\!}\!|^{2}\Bigr)\,,
    \label{eq:quantum fisher info pure states}
\end{align}
where $\ket{\dot{\psi}(\theta)\!}=\tfrac{\partial}{\partial\theta}\ket{\psi(\theta)\!}$. When some of the modes are not accessible and are traced out, Eq.~(\ref{eq:quantum fisher info pure states}) provides an upper bound on the QFI. To allow for more specific statements about the QFI, additional information about the initial states or transformations encoding~$\theta$ is required.

\begin{figure}[ht!]
\includegraphics[width=0.47\textwidth]{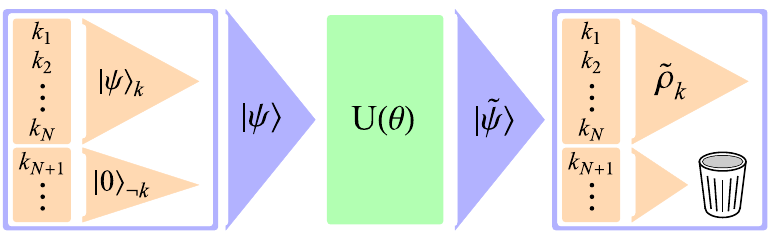}
\caption{\label{fig:metrology scheme}(Color online)
\textbf{Parameter estimation scheme}. To estimate the parameter~$\theta$,~a set of modes $k=\{k_{1},k_{2},\ldots,k_{N}\}$ is prepared in some initial states~$\ket{\psi}_{k}$, while all other modes are left in the vacuum. The unitary transformation~$U(\theta)$ encodes the parameter in the final state~$\ket{\tilde{\psi}}$, but measurements to determine~$\theta$ can only be performed on the marginal~$\tilde{\rho}_{k}$, while information in other modes is lost.}
\end{figure}

\section{QFI in the perturbative regime}

\subsection{Parameter estimation for pure states}

We now consider cases where $\theta$ is very small and close to zero. For this case the
transformation $U(\theta)$ is amenable to a perturbative approach and has a Taylor-Maclaurin expansion of the form
\begin{align}
    U(\theta)   &=\,U^{\raisebox{-0.5pt}{\tiny{(0)}}}\,+\,\theta\,U^{\raisebox{-0.5pt}{\tiny{(1)}}}\,+\,\theta^{2}\,U^{\raisebox{-0.5pt}{\tiny{(2)}}}\,+\,\mathcal{O}(\theta^{3})\,,
\end{align}
where $U^{\raisebox{-0.5pt}{\tiny{(0)}}}$ is~a unitary that encodes the free time evolution of the noninteracting
bosonic modes, while for $i>0$ the $U^{\raisebox{-0.5pt}{\tiny{(i)}}}$ represent (nonunitary) perturbations.
Consequently, the final state can be written as
\begin{align}
    \ket{\tilde{\psi}}  &=\,\ket{\tilde{\psi}^{\raisebox{-0.5pt}{\tiny{(0)}}}\!}\,+\,\theta\,\ket{\tilde{\psi}^{\raisebox{-0.5pt}{\tiny{(1)}}}\!}\,+\,
    \theta^{2}\,\ket{\tilde{\psi}^{\raisebox{-0.5pt}{\tiny{(2)}}}\!}\,+\,\mathcal{O}(\theta^{3})\,,
    \label{eq:pure state expansion}
\end{align}
where $\ket{\tilde{\psi}^{\raisebox{-0.5pt}{\tiny{(i)}}}\!}=U^{\raisebox{-0.5pt}{\tiny{(i)}}}\ket{\psi}$\ ($i=0,1,2,\ldots$). Note that $\ket{\tilde{\psi}}$ and $\ket{\tilde{\psi}^{\raisebox{-0.5pt}{\tiny{(0)}}}\!}$ are normalized, but the vectors $\ket{\tilde{\psi}^{\raisebox{-0.5pt}{\tiny{(1)}}}\!}$ $(i>0)$ generally are not. Substituting~(\ref{eq:pure state expansion}) into the expression for the QFI in~(\ref{eq:quantum fisher info pure states}), we arrive at
\begin{align}
    \mathcal{I}(\ket{\tilde{\psi}}) &=\,4\Bigl(\scpr{\!\tilde{\psi}^{\raisebox{-0.5pt}{\tiny{(1)}}}}{\tilde{\psi}^{\raisebox{-0.5pt}{\tiny{(1)}}}\!}\,-\,
    |\!\scpr{\!\tilde{\psi}^{\raisebox{-0.5pt}{\tiny{(0)}}}}{\tilde{\psi}^{\raisebox{-0.5pt}{\tiny{(1)}}}\!}\!|^{2}\Bigr)\,+\,\mathcal{O}(\theta)\,.
    \label{eq:quantum fisher info pure states perturbative}
\end{align}
Hence, given that all the bosonic modes are accessible, the QFI can be straightforwardly computed for any initial
(pure) state from just its linear perturbations.

As all our results utilize the perturbative approach,~a few important remarks concerning the applicability of the
perturbative approach are in order. In the latter we assume that all higher-order terms in~$\theta$
are small and are therefore neglected. To ensure such reasoning is justified, we regard our
approach as valid so long as the perturbation to the state $\ket{\tilde{\psi}^{\raisebox{-0.5pt}{\tiny{(0)}}}\!}$
remains small, such that $1-\mathcal{F}(\ket{\tilde{\psi}^{\raisebox{-0.5pt}{\tiny{(0)}}}\!},\ket{\tilde{\psi}})\ll1$. This
implies that $\theta^{2}\mathcal{I}(\ket{\psi})/4\ll1$ and all optimizations will be performed with this
constraint in mind.

\subsection{Tracing losses}

Thus far we have assumed that all bosonic modes are accessible and can be controlled. This is an unrealistic assumption
in practice, as only~a finite subset $k=\{k_{1},k_{2},\ldots,k_{N}\}$ of the modes can be addressed simultaneously.
Moreover, as the estimation strategy is to be optimized at~a fixed investment of energy,
it appears unwise to initially populate modes in the complementary subset $\lnot k$, which cannot be measured. We therefore assume that the preparation of
nontrivial initial states is also limited to~$k$. Hence, the input states that we consider are of the form $\ket{\psi}=
\ket{\psi}_{k}\ket{0}_{\lnot k}$. The unitary~$U(\theta)$, on the other hand, acts on all modes such that the final state
is~$\ket{\tilde{\psi}}=U(\theta)\ket{\psi}$, but only the reduced state $\tilde{\rho}_{k}=\tr_{\lnot k}\ket{\tilde{\psi}}\!
\bra{\tilde{\psi}}$ is accessible for the estimation of~$\theta$ (see Fig.~\ref{fig:metrology scheme}).
Thus, Eq.~(\ref{eq:quantum fisher info pure states perturbative}) provides an upper bound, i.e., $\mathcal{I}(\tilde{\rho}
_{k}(\theta))\leq\mathcal{I}(\ket{\tilde{\psi}(\theta)\!})$ for the precision with which~$\theta$ can be estimated.
However,~a precise expression for the losses incurred by tracing can be established as we will show now. Let us expand the reduced state $\tilde{\rho}_{k}(\theta)$ in powers of~$\theta$,
\begin{align}
    \tilde{\rho}_{k}(\theta)    &=\,\tilde{\rho}^{\raisebox{-0.5pt}{\tiny{(0)}}}_{k}\,+\,\theta\,\tilde{\rho}^{\raisebox{-0.5pt}{\tiny{(1)}}}_{k}\,+\,
    \theta^{2}\,\tilde{\rho}^{\raisebox{-0.5pt}{\tiny{(2)}}}_{k}\,+\,\mathcal{O}(\theta^{3})\,,
    \label{eq:mixed state expansion}
\end{align}
where the leading order is $\tilde{\rho}_{k}^{\raisebox{-0.5pt}{\tiny{(0)}}}=\tr_{\lnot k}\bigl(\ket{\tilde{\psi}^{\raisebox{-0.5pt}{\tiny{(0)}}}\!}\!\bra{\!\tilde{\psi}^{\raisebox{-0.5pt}{\tiny{(0)}}}\!}\bigr)$ and the corrections are given by
\begin{align}
    \tilde{\rho}_{k}^{\raisebox{-0.5pt}{\tiny{(1)}}}    &=
        \tr_{\lnot k}\bigl(\ket{\tilde{\psi}^{\raisebox{-0.5pt}{\tiny{(1)}}}\!}\!\bra{\!\tilde{\psi}^{\raisebox{-0.5pt}{\tiny{(0)}}}\!}+
        \ket{\tilde{\psi}^{\raisebox{-0.5pt}{\tiny{(0)}}}\!}\!\bra{\!\tilde{\psi}^{\raisebox{-0.5pt}{\tiny{(1)}}}\!}\bigr)\,,
        \label{eq:rho tilde k one}\\[1mm]
    \hspace{-2mm}\tilde{\rho}_{k}^{\raisebox{-0.5pt}{\tiny{(2)}}}    &=\tr_{\lnot k}\bigl(
    \ket{\tilde{\psi}^{\raisebox{-0.5pt}{\tiny{(0)}}}\!}\!\bra{\!\tilde{\psi}^{\raisebox{-0.5pt}{\tiny{(2)}}}\!}
    +\ket{\tilde{\psi}^{\raisebox{-0.5pt}{\tiny{(2)}}}\!}\!\bra{\!\tilde{\psi}^{\raisebox{-0.5pt}{\tiny{(0)}}}\!}
    +\ket{\tilde{\psi}^{\raisebox{-0.5pt}{\tiny{(1)}}}\!}\!\bra{\!\tilde{\psi}^{\raisebox{-0.5pt}{\tiny{(1)}}}\!}\bigr)\,.
    \label{eq:rho tilde k two}
\end{align}
As the modes are noninteracting, the free evolution $U^{\raisebox{-0.5pt}{\tiny{(0)}}}$ is~a local operation that leaves the vacuum invariant and we may write $\ket{\tilde{\psi}^{\raisebox{-0.5pt}{\tiny{(0)}}}\!}=U^{\raisebox{-0.5pt}{\tiny{(0)}}}\ket{\psi}_{k}\ket{0}_{\lnot k}=\ket{\tilde{\psi}^{\raisebox{-0.5pt}{\tiny{(0)}}}\!}_{k}\ket{0}_{\lnot k}$. It then easily follows that $\tilde{\rho}^{\raisebox{-0.5pt}{\tiny{(0)}}}_{k}=
\ket{\tilde{\psi}^{\raisebox{-0.5pt}{\tiny{(0)}}}\!}_{\hspace*{-0.5pt}kk\hspace*{-1pt}}\!
\bra{\tilde{\psi}^{\raisebox{-0.5pt}{\tiny{(0)}}}\!}$ is~a pure state. In~a similar way we may expand $\tilde{\rho}_{k}(\theta+d\theta)$ as
\begin{align}
    \tilde{\rho}_{k}(\theta+d\theta)    &=\tilde{\rho}_{k}(\theta)+d\theta\frac{\partial\tilde{\rho}_{k}(\theta)}{\partial\theta}
    +\frac{d\theta^{2}}{2}\,\frac{\partial^{2}\tilde{\rho}_{k}(\theta)}{\partial\theta^{2}}+\mathcal{O}(d\theta^{3})\nonumber\\
    &=\,\tilde{\rho}^{\raisebox{-0.5pt}{\tiny{(0)}}}_{k}\,+\,d\theta\,\tilde{\rho}^{\raisebox{-0.5pt}{\tiny{(1)}}}_{k}\,+\,
    d\theta^{2}\,\tilde{\rho}^{\raisebox{-0.5pt}{\tiny{(2)}}}_{k}\,+\,\mathcal{O}(\theta,d\theta^{3})\,.
    \label{eq:mixed state expansion dtheta}
\end{align}
Inserting~(\ref{eq:mixed state expansion}) and~(\ref{eq:mixed state expansion dtheta}) into Eq.~(\ref{eq:quantum fisher info definition}) and taking into account the unitarity of the transformation, we arrive at the central result
\begin{align}
    \mathcal{I}\bigl(\tilde{\rho}_{k}(\theta)\bigr)   &=\,-4\,_{\raisebox{-1pt}{\scriptsize{$k$}}\!}\!\bra{\tilde{\psi}^{\raisebox{-0.5pt}{\tiny{(0)}}}\!}
    \tilde{\rho}_{k}^{\raisebox{-0.5pt}{\tiny{(2)}}}
    \ket{\tilde{\psi}^{\raisebox{-0.5pt}{\tiny{(0)}}}\!}_{\!k}\,+\,\mathcal{O}(\theta)\,,
    \label{eq:quantum fisher info mixed states perturbative}
\end{align}
for which a detailed proof is given in the \hyperref[sec:appendix]{Appendix}.
We pause briefly to highlight the simplicity of Eq.~\eqref{eq:quantum fisher info mixed states perturbative}: One only
requires a \emph{single matrix element} in the second-order expansion of $\tilde{\rho}_k(\theta)$ in order to compute
the QFI, to leading order in $\theta$, for any input state~$\ket{\psi}_{k}$.

To compare Eq.~\eqref{eq:quantum fisher info mixed states perturbative} to the case where no information is lost to tracing, we decompose $\tilde{\rho}_{k}^{\raisebox{-0.5pt}{\tiny{(2)}}}$ as in Eq.~(\ref{eq:rho tilde k two}). Exploiting the unitarity of the transformation, one obtains (see the \hyperref[sec:appendix]{Appendix})
\begin{align}
    \mathcal{I}\bigl(\tilde{\rho}_{k}(\theta)\bigr)   &=\,\mathcal{I}(\ket{\tilde{\psi}})\,-\,4\sum\limits_{i\neq0}
    |\!\,_{\raisebox{-1pt}{\scriptsize{$k$}}\!}\!\bra{\tilde{\psi}^{\raisebox{-0.5pt}{\tiny{(0)}}}\!}
    \!\!\,_{\raisebox{-1pt}{\scriptsize{$\lnot k$}}\!}\!\bra{i}U^{\raisebox{-0.5pt}{\tiny{(1)}}}\!\ket{\psi}\!|^{2}+\mathcal{O}(\theta)
    \nonumber\\[-1mm]
    &=\,\mathcal{I}(\ket{\tilde{\psi}})\,-\,\Delta_{\mathrm{tr}}(\ket{\psi}_{k})\,+\,\mathcal{O}(\theta)\,,
    \label{eq:tracing losses}
\end{align}
where $\mathcal{I}(\ket{\tilde{\psi}})$ is given as in Eq.~(\ref{eq:quantum fisher info pure states perturbative}) and the
sum runs over all basis states of the Fock space of $\lnot k$ that are orthogonal to the vacuum state~$\ket{0}_{\!\lnot
k}$.  Note that Eq.~\eqref{eq:tracing losses} suggests that in order to compute the QFI to leading-order in $\theta$
one requires only leading order corrections to the initial state, whereas
Eq.~\eqref{eq:quantum fisher info mixed states perturbative} above implied that second-order corrections are needed.
However, as is shown in detail in the \hyperref[sec:appendix]{Appendix}, the requirement that the transformation is
unitary implies that the second-order correction in Eq.~\eqref{eq:quantum fisher info mixed states perturbative} can be
expressed in terms of the linear order corrections to the state.

\subsection{Parameter estimation from Bogoliubov transformations}

For the remainder of this paper, we will work with Eq.~(\ref{eq:tracing losses}) to investigate how tracing losses can be minimized and optimal scaling can be achieved. For this task, we will specialize our discussion to Bogoliubov transformations~\cite{BirrellDavies:QFbook}, that is, unitary transformations on the Fock space that can be viewed as linear transformations of the mode operators, i.e.,
\vspace*{-1.5mm}
\begin{align}
    a_{m}   &\mapsto\, \tilde{a}_{m}=\sum\limits_{n}\bigl(\alpha^{*}_{mn}\,a_{n}\,-\,\beta^{\,*}_{mn}\,a^{\dagger}_{n}\bigr)\,,
    \label{eq:Bogo transform}
\end{align}
where unitarity places restrictions on the complex coefficients~$\alpha_{mn}$ and~$\beta_{mn}$. Another way to view such transformations is as unitaries that are generated by quadratic combinations of the mode operators. In the perturbative regime that we consider here, the Bogoliubov coefficients are expanded as
\begin{align}
    \alphamn{mn}  &=  \alphahmn{0}{mn}\,+\,\alphahmn{1}{mn}\,\theta\,+\,\alphahmn{2}{mn}\,\theta^{2}\,+\,O(\theta^{3})\,,
    \label{eq:alphas small h expansion}\\
    \betamn{mn}   &=  \betahmn{1}{mn}\,\theta\,+\,\betahmn{2}{mn}\,\theta^{2}\,+\,O(\theta^{3})\,,
    \label{eq:betas small h expansion}
\end{align}
where $\alphahmn{0}{mn}=\delta_{mn}G_{n}=\delta_{mn}\exp(i\phi_{n})$ encodes the phases~$\phi_{n}$ that are
accumulated due to the free time evolution~$U^{\raisebox{-0.5pt}{\tiny{(0)}}}$. The coefficients $\alphamn{mn}$ and~
$\betamn{mn}$ are responsible for the shifts of single excitations between modes~$m$ and~$n$ or the creation and
annihilation of pairs of particles in these modes, respectively. Any initial state of the Fock space may simply be written
in terms of linear combinations of products of creation operators acting on the vacuum. The operators transform according
to Eq.~(\ref{eq:Bogo transform}), whereas the vacuum transforms as
\begin{align}
    \ket{0} &\mapsto\,\ket{0}\,-\theta\tfrac{1}{2}\sum_{p,q}\Gpstar{q}\betahmnstar{1}{pq}a_{p}^{\dagger}a_{q}^{\dagger}\ket{0}\,+\,\mathcal{O}(\theta^{2})\,,
    \label{eq:vacuum transformation}
\end{align}
where $\Gpstar{q}\betahmnstar{1}{pq}=\Gpstar{p}\betahmnstar{1}{qp}$ and $\betahmn{1}{pq}$ must be~a Hilbert-Schmidt operator $\sum_{p,q}|\betahmn{1}{pq}|<\infty$ to guarantee unitarity (see, e.g., Ref.~\cite[pp.~93 and 101]{Friis:PhD2013}). We now determine the tracing losses $\Delta_{\mathrm{tr}}(\ket{\psi}_{\!k})$ and the scaling of
$\mathcal{I}(\ket{\tilde{\psi}})$ with the average energy of the initial state in the above scenario.

The linear order operation $U^{\raisebox{-0.5pt}{\tiny{(1)}}}$ can be (uniquely) separated into (all) terms that leave the particle content of $k$ or $\lnot k$ invariant and~a term that correlates both sets via the creation or exchange of excitations, i.e.,
\vspace*{-1.0mm}
\begin{align}
    U^{\raisebox{-0.5pt}{\tiny{(1)}}}   &=\,U^{\raisebox{-0.5pt}{\tiny{(1)}}}_{k}\!\otimes U^{\raisebox{-0.5pt}{\tiny{(0)}}}_{\lnot k}
    \,+\,U^{\raisebox{-0.5pt}{\tiny{(0)}}}_{k}\!\otimes U^{\raisebox{-0.5pt}{\tiny{(1)}}}_{\lnot k}+U^{\raisebox{-0.5pt}{\tiny{(1)}}}_{k,\lnot k}\,.
    \label{eq:unitary linear correction splitting}
\end{align}
From Eq.~(\ref{eq:tracing losses}) it is obvious that $U^{\raisebox{-0.5pt}{\tiny{(1)}}}_{k}\!\!\otimes\!U^{\raisebox{-0.5pt}{\tiny{(0)}}}_{\lnot k}$ does not contribute to~$\Delta_{\mathrm{tr}}(\ket{\psi}_{\!k})$. The other two terms generally provide nonzero contributions. However, to linear order, the operation $U^{\raisebox{-0.5pt}{\tiny{(1)}}}_{\lnot k}$ populates~$\ket{0}_{\!\lnot k}$ with pairs of particles, whereas $U^{\raisebox{-0.5pt}{\tiny{(1)}}}_{k,\lnot k}$ may either create one particle each in~$k$ and~$\lnot k$ or shift one excitation from $k$ to $\lnot k$.
As a result, for the same fixed state $\ket{i}_{\!\lnot k}$, which contains at most two excitations, $\!\,_{\raisebox{-1pt}{\scriptsize{$\lnot k$}}\!}\!\bra{i}U^{\raisebox{-0.5pt}{\tiny{(1)}}}_{\lnot k}\!\ket{0}_{\!\lnot k}$ and $\,_{\raisebox{-1pt}{\scriptsize{$k$}}\!}\!\bra{\tilde{\psi}^{\raisebox{-0.5pt}{\tiny{(0)}}}\!}\!\!\,_{\raisebox{-1pt}{\scriptsize{$\lnot k$}}\!}\!\bra{i}U^{\raisebox{-0.5pt}{\tiny{(1)}}}_{k,\lnot k}\!\ket{\psi}$ cannot both be nonzero and hence contribute separately to the tracing loss. The latter term can be avoided by choosing~$\ket{\psi}_{\!k}$ such that superpositions of states that differ only by one excitation are excluded. The former term, on the other hand, cannot be avoided as it is independent of the state $\ket{\psi}_{\!k}$. It can, however, straightforwardly be computed using the vacuum state of Eq.~(\ref{eq:vacuum transformation}), for which $\,_{\raisebox{-1pt}{\scriptsize{$k$}}\!}\!\bra{0}\!\!\,_{\raisebox{-1pt}{\scriptsize{$\lnot k$}}\!}\!\bra{i}
U^{\raisebox{-0.5pt}{\tiny{(1)}}}_{k,\lnot k}\!\ket{0}_{\!k}\!\ket{0}_{\!\lnot k}=0$. For this case we find (see the \hyperref[sec:appendix]{Appendix})
\vspace*{-1.0mm}
\begin{align}
    \Delta_{\mathrm{tr}}(\ket{\psi}_{\!k})  &\geq\,\Delta_{\mathrm{tr}}(\ket{0}_{\!k})\,=\,2\sum\limits_{p,q\notin k}|\betahmn{1}{pq}|^{2}\,.
    \label{eq:tracing loss lower bound}
\end{align}
Thus, by choosing our initial state $\ket{\psi}_{\!k}$ such that superpositions of states that differ only by one
excitation are excluded, the inequality in Eq.~\eqref{eq:tracing loss lower bound} becomes a strict equality and the
QFI can be precisely determined to linear order in a straightforward way using Eq.~\eqref{eq:tracing losses}. We will restrict our discussion to such states from now on.

\subsection{Optimality}
We now show that the optimally achievable precision in this setup is Heisenberg scaling and how it can be realized even with nonentangled states. To determine the optimal states, the exact values of the Bogoliubov coefficients $\alphamn{mn}$
and $\betamn{mn}$ have to be known. This makes the procedure highly task specific. Nonetheless, it is possible to
identify families of states that scale optimally with the input energy under some minimal assumptions about the
Bogoliubov coefficients. First, consider a single-mode Fock state with~$n$ excitations. Using Eqs.~(\ref{eq:Bogo transform})-(\ref{eq:vacuum transformation}), this state transforms to
\begin{align}
    &\ket{\hspace*{-1pt}n_{k}\hspace*{-1pt}} \mapsto G_{k}^{\,n}\ket{\hspace*{-1pt}n_{k}\hspace*{-1pt}}
        -\theta \hspace*{1pt}G_{k}^{\,n}\Bigl[
        \tfrac{1}{2}\sqrt{n(n-1)}\,\Gpstar{k}\,\betahmn{1}{kk}\ket{\hspace*{-1pt}n\!-\!2_{k}\hspace*{-1pt}}
        \nonumber\\
    &\ -n\hspace*{1pt}\Gpstar{k}\alphahmn{1}{kk}\ket{\hspace*{-1pt}n_{k}\hspace*{-1pt}}
        +\tfrac{1}{2}\sqrt{(n+1)(n+2)}\,\Gpstar{k}\,\betahmnstar{1}{kk}\ket{\hspace*{-1pt}n\!+\!2_{k}\hspace*{-1pt}}
        \nonumber\\[1mm]
    &-\!\sum\limits_{p\neq k}\Bigl(\!
        \sqrt{n}\hspace*{1pt}\Gpstar{k}\alphahmn{1}{kp}\!\ket{\hspace*{-1pt}n\!-\!1_{k}\hspace*{-1pt}}\!\ket{\hspace*{-1pt}1_{p}\!}
        -\sqrt{n\!+\!1}\,\Gpstar{k}\betahmnstar{1}{pk}\!\ket{\hspace*{-1pt}n\!+\!1_{k}\hspace*{-1pt}}\!\ket{\hspace*{-1pt}1_{p}\!}
        \!\Bigr)\nonumber\\[-1mm]
    &+\tfrac{1}{2}\!\sum_{\substack{p,q\neq k\\ p\neq q}}
        \,\Gpstar{q}\,\betahmnstar{1}{pq}\ket{\hspace*{-1pt}n_{k}\hspace*{-1pt}}\!\ket{\hspace*{-1pt}1_{p}\!}\!\ket{\hspace*{-1pt}1_{q}\!}
        +\tfrac{1}{\sqrt{2}}\!\sum_{p\neq k}
        \,\Gpstar{p}\,\betahmnstar{1}{pp}\ket{\hspace*{-1pt}n_{k}\hspace*{-1pt}}\!\ket{\hspace*{-1pt}2_{p}\!}
        \Bigr]\nonumber\\[-2mm]
    &\qquad\qquad\,+\,\mathcal{O}(\theta^{2})\,.
    \label{eq:n particle transformation}
\end{align}
As $|\alphahmn{1}{kp}|=|\alphahmn{1}{pk}|$, one immediately obtains the QFI
\begin{align}
    \mathcal{I}(\ket{\hspace*{-1pt}n_{k}\hspace*{-1pt}})    &=\,
    2n(n+1)|\betahmn{1}{kk}|^{2}+4n\sum\limits_{p\neq k}\bigl(|\alphahmn{1}{pk}|^{2}+|\betahmn{1}{pk}|^{2}\bigr)\nonumber\\[-2mm]
    &\ \ +\,\mathcal{I}(\ket{0})\,+\,\mathcal{O}(\theta).
    \label{eq:quantum fisher info n particles}
\end{align}
Notice that, here, the coefficient $\alphahmn{1}{kk}$, which leaves the occupation number of the mode~$k$ unchanged, does not contribute to the QFI, but it may do so for superpositions of different particle numbers. However, when $\betahmn{1}{kk}\neq0$, which corresponds to single-mode squeezing transformations, terms that shift the population of the mode~$k$ by two excitations yield QFI that scales quadratically with~$n$, i.e., Heisenberg scaling. On the other hand, classical states, which in the present context are all coherent states, yield QFI that scales linearly with the average particle number~\cite{AhmadiBruschiFuentes2014}.

As~an example for the application of Eq.~(\ref{eq:quantum fisher info n particles}), consider~a scenario in quantum optics~\cite{MilburnChenJones1994} where only~a single mode labeled~$k$ is subject to weak single-mode squeezing, while all other modes are left invariant. The corresponding Bogoliubov transformation is represented by the coefficients $\alphamn{kk}=\cosh(\theta)=1+\tfrac{1}{2}\theta^{2}+\mathcal{O}(\theta^{3})$ and $\betamn{kk}=\sinh(\theta)=\theta+\mathcal{O}(\theta^{2})$. The optimal state for the estimation of~$\theta\ll1$ is hence already $\ket{\hspace*{-1pt}n_{k}\hspace*{-1pt}}$ and the QFI is $\mathcal{I}(\ket{\hspace*{-1pt}n_{k}\hspace*{-1pt}})=2n(n+1)+\mathcal{O}(\theta)$.

When the diagonal linear coefficients $\betahmn{1}{kk}$ vanish, as in the transformation of field modes of nonuniformly accelerating rigid cavities~\cite{BruschiFuentesLouko2012,FriisLeeLouko2013}, excitations in~a single mode are not sufficient to obtain Heisenberg scaling and one requires initial states with at least two occupied modes. For instance, when the modes $k$ and $k\pr$ can be controlled and $\betahmn{1}{kk}=\betahmn{1}{k\pr\! k\pr}=0$, the QFI for the state $\ket{\hspace*{-1pt}n_{k}\hspace*{-1pt}}\!\ket{\hspace*{-1pt}m_{k\pr}\hspace*{-2pt}}$ is (see the \hyperref[sec:appendix]{Appendix})
\begin{align}
    &\mathcal{I}(\ket{\hspace*{-1pt}n_{k}\hspace*{-1pt}}\!\ket{\hspace*{-1pt}m_{k\pr}\hspace*{-2pt}})    =
    8mn\bigl(|\alphahmn{1}{kk\pr}|^{2}+|\betahmn{1}{kk\pr}|^{2}\bigr)+4n\sum\limits_{p\neq k}\bigl(|\alphahmn{1}{pk}|^{2}
    \nonumber\\
    &+|\betahmn{1}{pk}|^{2}\bigr)+4m\!\sum\limits_{p\neq k\pr}\bigl(|\alphahmn{1}{pk\pr}|^{2}
    +|\betahmn{1}{pk\pr}|^{2}\bigr)+\mathcal{I}(\ket{0})+\mathcal{O}(\theta).\nonumber\\[-4mm]
    \label{eq:quantum fisher info m plus n particles}
\end{align}
The factor $mn$ allows the QFI to scale optimally with the average number of excitations for the estimation of beam-splitting ($\alphahmn{1}{kk\pr}$) and two-mode squeezing ($\betahmn{1}{kk\pr}$) terms. Indeed, inspection of the first-order expansion of $\ket{\hspace*{-1pt}n_{k}\hspace*{-1pt}}\!\ket{\hspace*{-1pt}m_{k\pr}\hspace*{-2pt}}$ reveals that Heisenberg scaling is the ultimate achievable precision scaling within our approach (see the \hyperref[sec:appendix]{Appendix}). For the explicit example of~a pure two-mode squeezing transformation, which may be realized in~a variety of physical systems, e.g., in superconducting circuitry~\cite{XueLiuSunNori2007}, the Bogoliubov coefficients are given by $\alphamn{kk}=\alphamn{k\pr\!k\pr}=\cosh(\theta)=1+\tfrac{1}{2}\theta^{2}+\mathcal{O}(\theta^{3})$ and $\betamn{kk\pr}=\betamn{k\pr\!k}=\sinh(\theta)=\theta+\mathcal{O}(\theta^{2})$. We hence recover the well known result that in this case the state $\ket{\hspace*{-1pt}n_{k}\hspace*{-1pt}}\!\ket{\hspace*{-1pt}n_{k\pr}\hspace*{-2pt}}$ is optimal and the QFI from Eq.~(\ref{eq:quantum fisher info m plus n particles}) yields
$\mathcal{I}(\ket{\hspace*{-1pt}n_{k}\hspace*{-1pt}}\!\ket{\hspace*{-1pt}n_{k\pr}\hspace*{-2pt}})=8n(n+1)+\mathcal{O}(\theta)$.

Whereas the scaling with respect to the average energy cannot be better than quadratic, even for arbitrary Gaussian transformations beyond the examples presented so far, one can still improve the constant prefactor in this scaling. For example, assuming $\betahmn{1}{kk}=0$ and $\alphahmn{1}{kk\pr},\betahmn{1}{kk\pr}\neq0$, the particle numbers ~$n$ and~$n\pm2$ in the state $\bigl(\ket{\hspace*{-1pt}n_{k}\hspace*{-1pt}}\!\ket{\hspace*{-1pt}n_{k\pr}\hspace*{-2pt}}
+\ket{\hspace*{-1pt}n_{k}\hspace*{-1pt}}\!\ket{\hspace*{-1pt}n\!-\!2_{k\pr}\hspace*{-2pt}}+\ket{\hspace*{-1pt}n_{k}\hspace*{-1pt}}\!\ket{\hspace*{-1pt}n\!+\!2_{k\pr}\hspace*{-2pt}}\bigr)/\sqrt{3}$ guarantee additional optimally scaling terms in the QFI while avoiding losses originating from $|\!\scpr{\!\tilde{\psi}^{\raisebox{-0.5pt}{\tiny{(0)}}}}{\tilde{\psi}^{\raisebox{-0.5pt}{\tiny{(1)}}}\!}\!|^{2}$.  The latter is nonzero whenever $\ket{\psi}$ contains superpositions of Fock states that can be converted into each other by changes of at most two excitations, such as $\ket{\hspace*{-1pt}n_{k}\!}\!\ket{n\!+\!1_{k\pr}\hspace*{-2pt}}$ and $\ket{n\!+\!1_{k}\!}\!\ket{\hspace*{-1pt}n_{k\pr}\hspace*{-2pt}}$, or $\ket{\hspace*{-1pt}n_{k}\!}\!\ket{\hspace*{-1pt}n_{k\pr}\hspace*{-2pt}}$ and $\ket{n\!+\!1_{k}\!}\!\ket{n\!+\!1_{k\pr}\hspace*{-2pt}}$.

Notice that in all of the examples considered so far the optimal states are not entangled, but can be regarded as highly squeezed and hence nonclassical. The final state may become entangled due to the transformation, but, as seen from~(\ref{eq:n particle transformation}) and~(\ref{eq:quantum fisher info n particles}), optimal scaling may arise from terms (here $\betahmn{1}{kk}\ket{\hspace*{-1pt}n\!-\!2_{k}\hspace*{-1pt}}$) that do not produce entanglement. Indeed, entanglement of the initial state is not necessary for optimality (a feature that was also noted in a related context in~\cite{SahotaQuesada2015}), although one may find entangled states that do admit Heisenberg scaling in precision such as the state $\bigl(\ket{n\!+\!1_{k}\!}\!\ket{n\!-\!1_{k\pr}\hspace*{-2pt}}+\ket{n\!-\!1_{k}\!}\!\ket{n\!+\!1_{k\pr}\hspace*{-2pt}}\bigr)/\sqrt{2}$, which also has minimal tracing loss. For the entangled state above, homodyne measurements would be required for optimal scaling, whereas the measurements for the
previous examples need to be number resolving. The choice of input state and measurement is thus also~a matter of what is practical in~a given experimental setup.

\section{Conclusion}

We have shown that the ultimate limits for estimating small parameters encoded in the linear transformations of the mode operators describing a set of noninteracting harmonic oscillators scale inversely proportional to the average input energy of the modes. Using~a perturbative approach for these Bogoliubov transformations, applicable when the parameter of interest is small, e.g., as in the case of nonuniformly moving cavities, we have derived analytical formulas for the QFI, which apply to initial overall pure states of all modes. We have provided~a lower bound on the losses when some of these modes cannot be controlled. Finally, we have identified the class of states that yield optimally scaling precision while exhibiting minimal tracing losses. This provides~a significant advancement beyond previous analysis of this problem~\cite{AhmadiBruschiFuentes2014}, which was limited to pure Gaussian states. An investigation of mixed initial states, such as the mixed Gaussian states that were investigated recently~\cite{SafranekLeeFuentes2015}, goes beyond the scope of the current paper, but is certainly of interest.

Our results open up the possibility to explore optimality in~a range of specific applications, where information about the Bogoliubov coefficients is available. Examples include field modes in non-uniformly moving cavities~\cite{BruschiFuentesLouko2012,FriisLeeLouko2013}, analog gravity phenomena~\cite{JaskulaPartridgeBonneauRuaudelBoironWestbrook2012}, and effects in superconducting materials~\cite{JohanssonJohanssonWilsonNori2010,LaehteenmaekiParaoanuHasselHakonen2013,DoukasLouko2014}.

\begin{acknowledgements}
We are grateful to Hans~J. Briegel, Vedran Dunjko, Antony R.~Lee, and Pavel Sekatski for valuable discussions and comments. This work was supported by the Austrian Science Fund (FWF) through Grants No.~SFB FoQuS F4012 and No.~P24273-N16.
\end{acknowledgements}


\appendix*
\section{Explicit calculations}\label{sec:appendix}
\renewcommand\appendixname{}
\renewcommand{\thesubsection}{A.\arabic{subsection}}

In this appendix we present explicit proofs of some key results presented in the main text. In particular, we give a detailed derivation of the QFI for the estimation of small parameters~$\theta$, encoded in Bogoliubov transformations, and we discuss the optimality of Heisenberg scaling.

\subsection{Reduced state QFI}\label{sec:reduced state QFI}

We start with the QFI $\mathcal{I}\bigl(\tilde{\rho}_{k}(\theta)\bigr)$ for~a mixed state $\tilde{\rho}_{k}(\theta)$, which we will show to be given by
\begin{align}
    \mathcal{I}\bigl(\tilde{\rho}_{k}(\theta)\bigr)   &=\,-4\,_{\raisebox{-1pt}{\scriptsize{$k$}}\!}\!\bra{\tilde{\psi}^{\raisebox{-0.5pt}{\tiny{(0)}}}\!}
    \tilde{\rho}_{k}^{\raisebox{-0.5pt}{\tiny{(2)}}}
    \ket{\tilde{\psi}^{\raisebox{-0.5pt}{\tiny{(0)}}}\!}_{\!k}\,+\,\mathcal{O}(\theta)\,,
    \label{eq:appendix quantum fisher info mixed states perturbative}
\end{align}
where $\tilde{\rho}_{k}(\theta)=\tr_{\lnot k}\ket{\tilde{\psi}}\!
\bra{\tilde{\psi}}$, $\ket{\tilde{\psi}}=U(\theta)\ket{\psi}$, and $U(\theta)$ have small-parameter expansions of the form
\begin{align}
    \tilde{\rho}_{k}(\theta)    &=\,\tilde{\rho}^{\raisebox{-0.5pt}{\tiny{(0)}}}_{k}\,+\,\theta\,\tilde{\rho}^{\raisebox{-0.5pt}{\tiny{(1)}}}_{k}\,+\,
    \theta^{2}\,\tilde{\rho}^{\raisebox{-0.5pt}{\tiny{(2)}}}_{k}\,+\,\mathcal{O}(\theta^{3})\,,
    \label{eq:appendix mixed state expansion}\\[1mm]
    \ket{\tilde{\psi}}  &=\,\ket{\tilde{\psi}^{\raisebox{-0.5pt}{\tiny{(0)}}}\!}\,+\,\theta\,\ket{\tilde{\psi}^{\raisebox{-0.5pt}{\tiny{(1)}}}\!}\,+\,
    \theta^{2}\,\ket{\tilde{\psi}^{\raisebox{-0.5pt}{\tiny{(2)}}}\!}\,+\,\mathcal{O}(\theta^{3})\,,
    \label{eq:appendix pure state expansion}\\[1mm]
    U(\theta)   &=\,U^{\raisebox{-0.5pt}{\tiny{(0)}}}\,+\,\theta\,U^{\raisebox{-0.5pt}{\tiny{(1)}}}\,+\,\theta^{2}\,U^{\raisebox{-0.5pt}{\tiny{(2)}}}\,+\,\mathcal{O}(\theta^{3})\,,
\end{align}
with $\ket{\tilde{\psi}^{\raisebox{-0.5pt}{\tiny{(0)}}}\!}=U^{\raisebox{-0.5pt}{\tiny{(0)}}}\ket{\psi}_{k}\ket{0}_{\lnot k}=\ket{\tilde{\psi}^{\raisebox{-0.5pt}{\tiny{(0)}}}\!}_{k}\ket{0}_{\lnot k}$.
The QFI of the state $\tilde{\rho}_{k}$ may be expressed via the Bures distance as
\begin{align}
    \mathcal{I}\bigl(\hspace*{-0.5pt}\tilde{\rho}_{k}(\theta)\hspace*{-0.5pt}\bigr) &=\,\lim_{d\theta\rightarrow0}8\frac{1-\sqrt{\mathcal{F}\bigl(\tilde{\rho}_{k}(\theta),\tilde{\rho}_{k}(\theta+d\theta)\bigr)}}{d\theta^{2}}\,,
    \label{eq:appendix quantum fisher info definition}
\end{align}
where the Uhlmann fidelity is given by
\begin{align}
    &\mathcal{F}\bigl(\tilde{\rho}_{k}(\theta),\tilde{\rho}_{k}(\theta+d\theta)\bigr)\\[1mm]
        &\qquad\equiv\Bigl(\tr\sqrt{\!\sqrt{\tilde{\rho}_{k}(\theta)}\,\tilde{\rho}_{k}(\theta+d\theta)\!\sqrt{\tilde{\rho}_{k}(\theta)}}\,\Bigr)^{2}.
\end{align}
With Eq.~(\ref{eq:appendix mixed state expansion}) at hand, the expansion in terms of $d\theta$ can be seen to be of the form
\begin{align}
    \tilde{\rho}_{k}(\theta+d\theta)    &=\tilde{\rho}_{k}(\theta)+d\theta\frac{\partial\tilde{\rho}_{k}(\theta)}{\partial\theta}
    +\frac{d\theta^{2}}{2}\,\frac{\partial^{2}\tilde{\rho}_{k}(\theta)}{\partial\theta^{2}}+\mathcal{O}(d\theta^{3})\nonumber\\[1mm]
    &=\,\tilde{\rho}^{\raisebox{-0.5pt}{\tiny{(0)}}}_{k}\,+\,d\theta\,\tilde{\rho}^{\raisebox{-0.5pt}{\tiny{(1)}}}_{k}\,+\,
    d\theta^{2}\,\tilde{\rho}^{\raisebox{-0.5pt}{\tiny{(2)}}}_{k}\,+\,\mathcal{O}(\theta,d\theta^{3})\,.
    \label{eq:appendix mixed state expansion dtheta}
\end{align}
Using Eqs.~(\ref{eq:appendix mixed state expansion}) and~(\ref{eq:appendix mixed state expansion dtheta}), we can then write
\begin{align}
    &\sqrt{\tilde{\rho}_{k}(\theta)}
    \tilde{\rho}_{k}(\theta+d\theta)
    \sqrt{\tilde{\rho}_{k}(\theta)}
    \nonumber\\[1mm]
    &=\,\sqrt{\tilde{\rho}_{k}^{\raisebox{-0.5pt}{\tiny{(0)}}}}
    \bigl(\tilde{\rho}^{\raisebox{-0.5pt}{\tiny{(0)}}}_{k}\,+\,d\theta\,\tilde{\rho}^{\raisebox{-0.5pt}{\tiny{(1)}}}_{k}\,+\,
    d\theta^{2}\,\tilde{\rho}^{\raisebox{-0.5pt}{\tiny{(2)}}}_{k}\bigr)
    \sqrt{\tilde{\rho}_{k}^{\raisebox{-0.5pt}{\tiny{(0)}}}}\,+\,\mathcal{O}(\theta,d\theta^{3})
    \nonumber\\[1mm]
    &=\,\ket{\tilde{\psi}^{\raisebox{-0.5pt}{\tiny{(0)}}}\!}_{\hspace*{-0.5pt}kk\hspace*{-1pt}}\!
    \bra{\tilde{\psi}^{\raisebox{-0.5pt}{\tiny{(0)}}}\!}\bigl(1\,+\,d\theta
    \,_{\raisebox{-1pt}{\scriptsize{$k$}}\!}\!\bra{\tilde{\psi}^{\raisebox{-0.5pt}{\tiny{(0)}}}\!}
    \tilde{\rho}_{k}^{\raisebox{-0.5pt}{\tiny{(1)}}}
    \ket{\tilde{\psi}^{\raisebox{-0.5pt}{\tiny{(0)}}}\!}_{\!k}\nonumber\\[1mm]
    &\ \ \ \ +\,d\theta^{2}\,_{\raisebox{-1pt}{\scriptsize{$k$}}\!}\!\bra{\tilde{\psi}^{\raisebox{-0.5pt}{\tiny{(0)}}}\!}
    \tilde{\rho}_{k}^{\raisebox{-0.5pt}{\tiny{(2)}}}
    \ket{\tilde{\psi}^{\raisebox{-0.5pt}{\tiny{(0)}}}\!}_{\!k}\bigr)\,+\,\mathcal{O}(\theta,d\theta^{3})\,,
    \label{eq:appendix fidelity proof 1}
\end{align}
where we have made use of the fact that $\tilde{\rho}^{\raisebox{-0.5pt}{\tiny{(0)}}}_{k}=
\ket{\tilde{\psi}^{\raisebox{-0.5pt}{\tiny{(0)}}}\!}_{\hspace*{-0.5pt}kk\hspace*{-1pt}}\!
\bra{\tilde{\psi}^{\raisebox{-0.5pt}{\tiny{(0)}}}\!}$ is pure. Now we can take~a closer look at the term linear in $d\theta$, i.e.,
\begin{align}
    & \,_{\raisebox{-1pt}{\scriptsize{$k$}}\!}\!\bra{\tilde{\psi}^{\raisebox{-0.5pt}{\tiny{(0)}}}\!}
    \tilde{\rho}_{k}^{\raisebox{-0.5pt}{\tiny{(1)}}}
    \ket{\tilde{\psi}^{\raisebox{-0.5pt}{\tiny{(0)}}}\!}_{\!k}\,\nonumber\\[1mm]
    &=\,\,_{\raisebox{-1pt}{\scriptsize{$k$}}\!}\!\bra{\tilde{\psi}^{\raisebox{-0.5pt}{\tiny{(0)}}}\!}
    \tr_{\lnot k}\bigl(\ket{\tilde{\psi}^{\raisebox{-0.5pt}{\tiny{(1)}}}\!}\!\bra{\tilde{\psi}^{\raisebox{-0.5pt}{\tiny{(0)}}}\!}+
    \mathrm{H.~c.}
    \bigr)
    \ket{\tilde{\psi}^{\raisebox{-0.5pt}{\tiny{(0)}}}\!}_{\!k}\nonumber\\[1mm]
    &=_{\raisebox{-1.5pt}{\scriptsize{$k$}}\!}\!\bra{\hspace*{-1pt}\tilde{\psi}^{\raisebox{-0.5pt}{\tiny{(0)}}}\!\hspace*{-1pt}}\!\sum_{i}\!\!\,_{\raisebox{-1pt}{\scriptsize{$\lnot k$}}\!}\!\bra{i}
        \!\hspace*{-1pt}\Bigl(\hspace*{-1pt}U^{\raisebox{-0.5pt}{\tiny{(1)}}}\!\ket{\hspace*{-1pt}\psi\hspace*{-1pt}}\!\hspace*{-1pt}
        \bigl(\!\,_{\raisebox{-1pt}{\scriptsize{$k$}}\!}\!\bra{\hspace*{-1pt}\tilde{\psi}^{\raisebox{-0.5pt}{\tiny{(0)}}}\!}\!\,_{\raisebox{-1pt}{\scriptsize{$\lnot k$}}\!}\!\bra{0}\bigr)
        \hspace*{-1pt}+\hspace*{-1pt}\mathrm{H.c.}\hspace*{-1pt}\Bigr)\hspace*{-1pt}\!\ket{i}_{\!\lnot k}\!\ket{\hspace*{-1pt}\tilde{\psi}^{\raisebox{-0.5pt}{\tiny{(0)}}}\!\hspace*{-1pt}}_{\!k}\nonumber\\[1mm]
    &=\,_{\raisebox{-1pt}{\scriptsize{$k$}}\!}\!\bra{\tilde{\psi}^{\raisebox{-0.5pt}{\tiny{(0)}}}\!}
        \!\,_{\raisebox{-1pt}{\scriptsize{$\lnot k$}}\!}\!\bra{0}U^{\raisebox{-0.5pt}{\tiny{(1)}}}\!\ket{\psi}
        \,+\,\bra{\psi}U^{\raisebox{-0.5pt}{\tiny{(1)$\dagger$}}}\ket{\tilde{\psi}^{\raisebox{-0.5pt}{\tiny{(0)}}}\!}_{\!k}\ket{0}_{\!\lnot k}\nonumber\\[1mm]
    &=\,\bra{\psi}U^{\raisebox{-0.5pt}{\tiny{(0)$\dagger$}}}U^{\raisebox{-0.5pt}{\tiny{(1)}}}\ket{\psi}\,+\,
        \bra{\psi}U^{\raisebox{-0.5pt}{\tiny{(1)$\dagger$}}}U^{\raisebox{-0.5pt}{\tiny{(0)}}}\ket{\psi}\,,
\end{align}
where we have used $\ket{\tilde{\psi}^{\raisebox{-0.5pt}{\tiny{(0)}}}\!}_{\!k}\ket{0}_{\!\lnot k}=\ket{\tilde{\psi}^{\raisebox{-0.5pt}{\tiny{(0)}}}\!}=U^{\raisebox{-0.5pt}{\tiny{(0)}}}\ket{\psi}$. On the other hand, the unitarity of the transformation requires
\begin{align}
    &\mathds{1}=U^{\dagger}U=U^{\raisebox{-0.5pt}{\tiny{(0)$\dagger$}}}U^{\raisebox{-0.5pt}{\tiny{(0)}}}+\theta
        \bigl(U^{\raisebox{-0.5pt}{\tiny{(0)$\dagger$}}}U^{\raisebox{-0.5pt}{\tiny{(1)}}}+U^{\raisebox{-0.5pt}{\tiny{(1)$\dagger$}}}U^{\raisebox{-0.5pt}{\tiny{(0)}}}\bigr)
        \nonumber\\[1mm]
        &+\theta^{2}
        \bigl(U^{\raisebox{-0.5pt}{\tiny{(0)$\dagger$}}}U^{\raisebox{-0.5pt}{\tiny{(2)}}}+U^{\raisebox{-0.5pt}{\tiny{(2)$\dagger$}}}U^{\raisebox{-0.5pt}{\tiny{(0)}}}
        +U^{\raisebox{-0.5pt}{\tiny{(1)$\dagger$}}}U^{\raisebox{-0.5pt}{\tiny{(1)}}}\bigr)
        +\mathcal{O}(\theta^{3})\,.
        \label{eq:appendix unitarity}
\end{align}
Since $U^{\raisebox{-0.5pt}{\tiny{(0)$\dagger$}}}U^{\raisebox{-0.5pt}{\tiny{(0)}}}=\mathds{1}$, it follows that all other terms on the right-hand side of Eq.~(\ref{eq:appendix unitarity}) must vanish. We hence find
\begin{align}
_{\raisebox{-1pt}{\scriptsize{$k$}}\!}\!\bra{\tilde{\psi}^{\raisebox{-0.5pt}{\tiny{(0)}}}\!}
    \tilde{\rho}_{k}^{\raisebox{-0.5pt}{\tiny{(1)}}}
    \ket{\tilde{\psi}^{\raisebox{-0.5pt}{\tiny{(0)}}}\!}_{\!k}  &=\,0\,.
\end{align}
Reinserting this into Eq.~(\ref{eq:appendix fidelity proof 1}), we obtain the Uhlmann fidelity
\begin{align}
    \mathcal{F}(\tilde{\rho}_{k}(\theta),\tilde{\rho}_{k}(\theta+d\theta))  &=\,1\,+\,d\theta^{2}\,\,_{\raisebox{-1pt}{\scriptsize{$k$}}\!}\!\bra{\tilde{\psi}^{\raisebox{-0.5pt}{\tiny{(0)}}}\!}
    \tilde{\rho}_{k}^{\raisebox{-0.5pt}{\tiny{(2)}}}
    \ket{\tilde{\psi}^{\raisebox{-0.5pt}{\tiny{(0)}}}\!}_{\!k}\nonumber\\[1mm]
    &\ \,+\,\mathcal{O}(\theta,d\theta^{3})\,,
\end{align}
which in turn yields the QFI of Eq.~(\ref{eq:appendix quantum fisher info mixed states perturbative}).
\qed\\

\subsection{Relation to pure state QFI}\label{sec:relation to pure state QFI}

Next we show that the QFI of the mixed state $\tilde{\rho}_{k}(\theta)$, arising from tracing out some of the modes from the pure state $\ket{\tilde{\psi}}$, that is, $\tilde{\rho}_{k}=\tr_{\lnot k}\ket{\tilde{\psi}}\!
\bra{\tilde{\psi}}$, may be related to the QFI of the latter pure state via the expression
\begin{align}
    \mathcal{I}\bigl(\tilde{\rho}_{k}(\theta)\bigr)   &=\,\mathcal{I}(\ket{\tilde{\psi}})\,-\,4\sum\limits_{i\neq0}
    |\!\,_{\raisebox{-1pt}{\scriptsize{$k$}}\!}\!\bra{\tilde{\psi}^{\raisebox{-0.5pt}{\tiny{(0)}}}\!}
    \!\!\,_{\raisebox{-1pt}{\scriptsize{$\lnot k$}}\!}\!\bra{i}U^{\raisebox{-0.5pt}{\tiny{(1)}}}\!\ket{\psi}\!|^{2}+\mathcal{O}(\theta)
    \nonumber\\[-1mm]
    &=\,\mathcal{I}(\ket{\tilde{\psi}})\,-\,\Delta_{\mathrm{tr}}(\ket{\psi}_{k})\,+\,\mathcal{O}(\theta)\,,
    \label{eq:appendix tracing losses}
\end{align}
where
\begin{align}
    \mathcal{I}(\ket{\tilde{\psi}}) &=\,4\Bigl(\scpr{\!\tilde{\psi}^{\raisebox{-0.5pt}{\tiny{(1)}}}}{\tilde{\psi}^{\raisebox{-0.5pt}{\tiny{(1)}}}\!}\,-\,
    |\!\scpr{\!\tilde{\psi}^{\raisebox{-0.5pt}{\tiny{(0)}}}}{\tilde{\psi}^{\raisebox{-0.5pt}{\tiny{(1)}}}\!}\!|^{2}\Bigr)\,+\,\mathcal{O}(\theta)\,.
    \label{eq:appendix quantum fisher info pure states perturbative}
\end{align}
Starting from Eqs.~(\ref{eq:appendix quantum fisher info mixed states perturbative}) and noting that $\tilde{\rho}_{k}^{\raisebox{-0.5pt}{\tiny{(2)}}}$ can be written as
\begin{align}
    \hspace{-2mm}\tilde{\rho}_{k}^{\raisebox{-0.5pt}{\tiny{(2)}}}    &=\tr_{\lnot k}\bigl(
    \ket{\tilde{\psi}^{\raisebox{-0.5pt}{\tiny{(0)}}}\!}\!\bra{\!\tilde{\psi}^{\raisebox{-0.5pt}{\tiny{(2)}}}\!}
    +\ket{\tilde{\psi}^{\raisebox{-0.5pt}{\tiny{(2)}}}\!}\!\bra{\!\tilde{\psi}^{\raisebox{-0.5pt}{\tiny{(0)}}}\!}
    +\ket{\tilde{\psi}^{\raisebox{-0.5pt}{\tiny{(1)}}}\!}\!\bra{\!\tilde{\psi}^{\raisebox{-0.5pt}{\tiny{(1)}}}\!}\bigr)\,,
    \label{eq:appendix rho tilde k two}
\end{align}
we compute
\begin{align}
    &\,_{\raisebox{-1pt}{\scriptsize{$k$}}\!}\!\bra{\tilde{\psi}^{\raisebox{-0.5pt}{\tiny{(0)}}}\!}
        \tilde{\rho}_{k}^{\raisebox{-0.5pt}{\tiny{(2)}}}
        \ket{\tilde{\psi}^{\raisebox{-0.5pt}{\tiny{(0)}}}\!}_{\!k}
        =\,_{\raisebox{-1pt}{\scriptsize{$k$}}\!}\!\bra{\tilde{\psi}^{\raisebox{-0.5pt}{\tiny{(0)}}}\!}\tr_{\lnot k}\Bigl[
        \ket{\tilde{\psi}^{\raisebox{-0.5pt}{\tiny{(0)}}}\!}\!\bra{\!\tilde{\psi}^{\raisebox{-0.5pt}{\tiny{(2)}}}\!}
    \nonumber\\[1mm]
    &\qquad   +\ket{\tilde{\psi}^{\raisebox{-0.5pt}{\tiny{(2)}}}\!}\!\bra{\!\tilde{\psi}^{\raisebox{-0.5pt}{\tiny{(0)}}}\!}
        +\ket{\tilde{\psi}^{\raisebox{-0.5pt}{\tiny{(1)}}}\!}\!\bra{\!\tilde{\psi}^{\raisebox{-0.5pt}{\tiny{(1)}}}\!}\Bigr]
        \ket{\tilde{\psi}^{\raisebox{-0.5pt}{\tiny{(0)}}}\!}_{\!k}
        \nonumber\\[1mm]
    &\qquad   =\,_{\raisebox{-1pt}{\scriptsize{$k$}}\!}\!\bra{\tilde{\psi}^{\raisebox{-0.5pt}{\tiny{(0)}}}\!}
    \sum_{i}\!\!\,_{\raisebox{-1pt}{\scriptsize{$\lnot k$}}\!}\!\bra{i}
        \!\Bigl[\Bigl(\ket{\tilde{\psi}^{\raisebox{-0.5pt}{\tiny{(0)}}}\!}_{\!k}\ket{0}_{\!\lnot k}\!\bra{\!\psi\!}
        U^{\raisebox{-0.5pt}{\tiny{(2)$\dagger$}}}+\mathrm{H.~c.}\Bigr)
        \nonumber\\
    &\qquad\qquad  +U^{\raisebox{-0.5pt}{\tiny{(1)}}}\ket{\!\psi\!}\!\bra{\!\psi\!}
        U^{\raisebox{-0.5pt}{\tiny{(1)$\dagger$}}}\Bigr]\ket{i}_{\!\lnot k}\ket{\tilde{\psi}^{\raisebox{-0.5pt}{\tiny{(0)}}}\!}_{\!k}
        \nonumber\\[1mm]
    &\qquad =\,\bra{\psi}U^{\raisebox{-0.5pt}{\tiny{(2)$\dagger$}}}U^{\raisebox{-0.5pt}{\tiny{(0)}}}
        +U^{\raisebox{-0.5pt}{\tiny{(0)$\dagger$}}}U^{\raisebox{-0.5pt}{\tiny{(2)}}}\ket{\psi}\nonumber\\[1mm]
    &\qquad\qquad  +\sum_{i}|\,_{\raisebox{-1pt}{\scriptsize{$k$}}\!}\!\bra{\tilde{\psi}^{\raisebox{-0.5pt}{\tiny{(0)}}}\!}
        \!\,_{\raisebox{-1pt}{\scriptsize{$\lnot k$}}\!}\!\bra{i}U^{\raisebox{-0.5pt}{\tiny{(1)}}}\ket{\psi}\!|^{2}\,.
        \label{eq:appendix rho tilde k two not proof 1}
\end{align}
For the first term we can use the unitarity of the transformation from Eq.~(\ref{eq:appendix unitarity}) such that
\begin{align}
    \bra{\psi}U^{\raisebox{-0.5pt}{\tiny{(2)$\dagger$}}}U^{\raisebox{-0.5pt}{\tiny{(0)}}}
        +U^{\raisebox{-0.5pt}{\tiny{(0)$\dagger$}}}U^{\raisebox{-0.5pt}{\tiny{(2)}}}\ket{\psi}  & = -\bra{\psi}U^{\raisebox{-0.5pt}{\tiny{(1)$\dagger$}}}U^{\raisebox{-0.5pt}{\tiny{(1)}}}\ket{\psi}
        \nonumber\\[1mm]
    &=-\scpr{\!\tilde{\psi}^{\raisebox{-0.5pt}{\tiny{(1)}}}}{\tilde{\psi}^{\raisebox{-0.5pt}{\tiny{(1)}}}\!}\,.
    \label{eq:appendix rho tilde k two not proof 2}
\end{align}
In the remaining part, the vacuum state $\ket{0}_{\!\lnot k}$ may be extracted from the sum such that
\begin{align}
    &\sum_{i}|\,_{\raisebox{-1pt}{\scriptsize{$k$}}\!}\!\bra{\tilde{\psi}^{\raisebox{-0.5pt}{\tiny{(0)}}}\!}
        \!\,_{\raisebox{-1pt}{\scriptsize{$\lnot k$}}\!}\!\bra{i}U^{\raisebox{-0.5pt}{\tiny{(1)}}}\ket{\psi}\!|^{2}
        =|\,_{\raisebox{-1pt}{\scriptsize{$k$}}\!}\!\bra{\tilde{\psi}^{\raisebox{-0.5pt}{\tiny{(0)}}}\!}
        \!\,_{\raisebox{-1pt}{\scriptsize{$\lnot k$}}\!}\!\bra{0}U^{\raisebox{-0.5pt}{\tiny{(1)}}}\ket{\psi}\!|^{2}
        \nonumber\\
    &\qquad +\sum_{i\neq 0}|\,_{\raisebox{-1pt}{\scriptsize{$k$}}\!}\!\bra{\tilde{\psi}^{\raisebox{-0.5pt}{\tiny{(0)}}}\!}
        \!\,_{\raisebox{-1pt}{\scriptsize{$\lnot k$}}\!}\!\bra{i}U^{\raisebox{-0.5pt}{\tiny{(1)}}}\ket{\psi}|^{2}
        \nonumber\\
        &=|\!\scpr{\!\tilde{\psi}^{\raisebox{-0.5pt}{\tiny{(0)}}}}{\tilde{\psi}^{\raisebox{-0.5pt}{\tiny{(1)}}}\!}\!|^{2}
        +\sum_{i\neq 0}|\,_{\raisebox{-1pt}{\scriptsize{$k$}}\!}\!\bra{\tilde{\psi}^{\raisebox{-0.5pt}{\tiny{(0)}}}\!}
        \!\,_{\raisebox{-1pt}{\scriptsize{$\lnot k$}}\!}\!\bra{i}U^{\raisebox{-0.5pt}{\tiny{(1)}}}\ket{\psi}\!|^{2}\,.
    \label{eq:appendix rho tilde k two not proof 3}
\end{align}
Combining~(\ref{eq:appendix rho tilde k two not proof 1})-(\ref{eq:appendix rho tilde k two not proof 3}) and comparing with Eq.~(\ref{eq:appendix quantum fisher info pure states perturbative}), we then immediately arrive at the result of Eq.~(\ref{eq:appendix tracing losses}).\qed

\subsection{Tracing Loss}\label{sec:tracing loss}

We now specialize to the case where $U(\theta)$ is~a Bogoliubov transformation that is represented as~a linear map of the mode operators, that is,
\begin{align}
    a_{m}   &\mapsto\, \tilde{a}_{m}=\sum\limits_{n}\bigl(\alpha^{*}_{mn}\,a_{n}\,-\,\beta^{\,*}_{mn}\,a^{\dagger}_{n}\bigr)\,,
    \label{eq:appendix Bogo transform}
\end{align}
where the complex coefficients $\alphamn{mn}$ and $\betamn{mn}$ can be expanded as
\begin{align}
    \alphamn{mn}  &=  \alphahmn{0}{mn}\,+\,\alphahmn{1}{mn}\,\theta\,+\,\alphahmn{2}{mn}\,\theta^{2}\,+\,O(\theta^{3})\,,
    \label{eq:appendix alphas small h expansion}\\
    \betamn{mn}   &=  \betahmn{1}{mn}\,\theta\,+\,\betahmn{2}{mn}\,\theta^{2}\,+\,O(\theta^{3})\,.
    \label{eq:appendix betas small h expansion}
\end{align}
In such~a scenario, the vacuum state can be shown (see, e.g., Ref.~\cite[p.~100]{Friis:PhD2013}) to transform as
\begin{align}
    \ket{0} &\mapsto\,\ket{0}\,-\theta\tfrac{1}{2}\sum_{p,q}\Gpstar{q}\betahmnstar{1}{pq}a_{p}^{\dagger}a_{q}^{\dagger}\ket{0}\,+\,\mathcal{O}(\theta^{2})\,,
    \label{eq:appendix vacuum transformation}
\end{align}
where $\Gpstar{q}\betahmnstar{1}{pq}=\Gpstar{p}\betahmnstar{1}{qp}$ due to unitarity (see, e.g., Ref.~\cite[pp.~93]{Friis:PhD2013}). To evaluate the tracing loss for the vacuum, for which this loss is minimal, note that the correction $U^{\raisebox{-0.5pt}{\tiny{(1)}}}$ can be uniquely split up into terms $U^{\raisebox{-0.5pt}{\tiny{(0)}}}_{k}\!\otimes U^{\raisebox{-0.5pt}{\tiny{(1)}}}_{\lnot k}$ and $U^{\raisebox{-0.5pt}{\tiny{(1)}}}_{k}\!\otimes U^{\raisebox{-0.5pt}{\tiny{(0)}}}_{\lnot k}$, which act trivially on the subspaces~$k$ or $\lnot k$, respectively, and the term $U^{\raisebox{-0.5pt}{\tiny{(1)}}}_{k,\lnot k}$, which correlates them, that is,
\begin{align}
    U^{\raisebox{-0.5pt}{\tiny{(1)}}}   &=\,U^{\raisebox{-0.5pt}{\tiny{(1)}}}_{k}\!\otimes U^{\raisebox{-0.5pt}{\tiny{(0)}}}_{\lnot k}
    \,+\,U^{\raisebox{-0.5pt}{\tiny{(0)}}}_{k}\!\otimes U^{\raisebox{-0.5pt}{\tiny{(1)}}}_{\lnot k}+U^{\raisebox{-0.5pt}{\tiny{(1)}}}_{k,\lnot k}\,.
    \label{eq:appendix unitary linear correction splitting}
\end{align}
For the initial vacuum state $\ket{\psi}_{\!k}=\ket{0}_{\!k}$, only the term $U^{\raisebox{-0.5pt}{\tiny{(0)}}}_{k}\!\otimes U^{\raisebox{-0.5pt}{\tiny{(1)}}}_{\lnot k}$ contributes to the tracing loss, which we identify from the vacuum transformation as
\begin{align}
    U^{\raisebox{-0.5pt}{\tiny{(0)}}}_{k}\!\otimes U^{\raisebox{-0.5pt}{\tiny{(1)}}}_{\lnot k}  &=\,
    -\tfrac{1}{2}\sum_{p,q\notin k}G_{q}^{*}\betahmnstar{1}{pq}a_{p}^{\dagger}a_{q}^{\dagger}\,.
    \label{eq:appendix vacuum transformation notk terms}
\end{align}
With this, we may write the tracing loss for the initial vacuum state as
\begin{align}
    &\Delta_{\mathrm{tr}}(\ket{0}_{\!k}) \,=\,4\sum\limits_{i\neq0}
    |
    \!\,_{\raisebox{-1pt}{\scriptsize{$k$}}\!}\!\bra{0}
    \!\,_{\raisebox{-1pt}{\scriptsize{$\lnot k$}}\!}\!\bra{i}
    U^{\raisebox{-0.5pt}{\tiny{(0)}}}_{k}\otimes U^{\raisebox{-0.5pt}{\tiny{(1)}}}_{\lnot k}\!\ket{0}_{\!k}\!\ket{0}_{\!\lnot k}
    |^{2}\nonumber\\
    &=\,
    2\!\sum_{p\pr\!,q\pr\notin k}|
    \tfrac{1}{2}\!\!\sum_{p,q\notin k}\!G_{q}^{*}\betahmnstar{1}{pq}
    \!\,_{\raisebox{-1pt}{\scriptsize{$\lnot k$}}\!}\!\bra{0}a_{p\pr}a_{q\pr}a_{p}^{\dagger}a_{q}^{\dagger}
    \ket{0}_{\!\lnot k}
    |^{2},
    \label{eq:appendix vacuum tracing loss 1}
\end{align}
where we note that, as opposed to the sum over~$i$, the sum over all $p\pr$ and $q\pr$ counts every basis vector twice and~a factor of~$\tfrac{1}{2}$ therefore has to be included in the conversion. Since unitarity requires $G_{q}^{*}\betahmn{1}{pq}=G_{p}^{*}\betahmn{1}{qp}$ (see Ref.~\cite[pp.~93]{Friis:PhD2013}) and $\bra{0}a_{p\pr}a_{q\pr}a_{p}^{\dagger}a_{q}^{\dagger}\ket{0}=\delta_{pp\pr}\delta_{qq\pr}+\delta_{pq\pr}\delta_{qp\pr}$ we arrive at
\begin{align}
    \Delta_{\mathrm{tr}}(\ket{0}_{\!k})\,=\,2\sum\limits_{p,q\notin k}|\betahmn{1}{pq}|^{2}\,\leq\,\Delta_{\mathrm{tr}}(\ket{\psi}_{\!k})\,.
    \label{eq:appendix vacuum tracing loss 2}
\end{align}

\subsection{Transformation \& QFI for Example States}\label{sec:transformation and QFI for example states}

We now give the explicit expression for the state transformation of $\ket{\hspace*{-1pt}n_{k}\hspace*{-1pt}}\!\ket{\hspace*{-1pt}m_{k\pr}\hspace*{-2pt}}$ and use the result to calculate the corresponding QFI.
From Eqs.~(\ref{eq:appendix Bogo transform})-(\ref{eq:appendix vacuum transformation}) we can write
\begin{align}
    &\ket{\hspace*{-1pt}n_{k}\hspace*{-1pt}}\!\ket{\hspace*{-1pt}m_{k\pr}\hspace*{-2pt}}    \,\mapsto\,
        G_{k}^{n}G_{k\pr}^{m}\ket{\hspace*{-1pt}n_{k}\hspace*{-1pt}}\!\ket{\hspace*{-1pt}m_{k\pr}\hspace*{-2pt}}
        +\theta\, G_{k}^{n}G_{k\pr}^{m}\Bigl[\nonumber\\[1mm]
    &+n\hspace*{1pt}\Gpstar{k}\alphahmn{1}{kk}\ket{\hspace*{-1pt}n_{k}\hspace*{-1pt}}\!\ket{\hspace*{-1pt}m_{k\pr}\hspace*{-2pt}}
        -\tfrac{1}{2}\sqrt{n(n\!-\!1)}\,\Gpstar{k}\,\betahmn{1}{kk}
        \ket{\hspace*{-1pt}n\!-\!2_{k}\hspace*{-1pt}}\!\ket{\hspace*{-1pt}m_{k\pr}\hspace*{-2pt}}\nonumber\\[1mm]
    &-\tfrac{1}{2}\sqrt{(n\!+\!1)(n\!+\!2)}\,\Gpstar{k}\,\betahmnstar{1}{kk}
        \ket{\hspace*{-1pt}n\!+\!2_{k}\hspace*{-1pt}}\!\ket{\hspace*{-1pt}m_{k\pr}\hspace*{-2pt}}\nonumber\\[1mm]
    &+m\hspace*{0pt}\Gpstar{k\pr}\alphahmn{1}{k\pr\!k\pr}\ket{\hspace*{-1pt}n_{k}\hspace*{-1pt}}\!\ket{\hspace*{-1pt}m_{k\pr}\hspace*{-2pt}}
        \!-\!\tfrac{1}{2}\hspace*{-1pt}\sqrt{\hspace*{-1pt}m(m\!-\!1)}\Gpstar{k\pr}\betahmn{1}{k\pr\!k\pr}
        \ket{\hspace*{-1pt}n_{k}\hspace*{-1pt}}\!\ket{\hspace*{-1pt}m\!-\!2_{k\pr}\hspace*{-2pt}}\nonumber\\[1mm]
    &-\tfrac{1}{2}\sqrt{(m\!+\!1)(m\!+\!2)}\,\Gpstar{k\pr}\,\betahmnstar{1}{k\pr\!k\pr}
        \ket{\hspace*{-1pt}n_{k}\hspace*{-1pt}}\!\ket{\hspace*{-1pt}m\!+\!2_{k\pr}\hspace*{-2pt}}\nonumber\\[1mm]
    &\ +\sqrt{m(n+1)}\,\Gpstar{k\pr}\alphahmn{1}{k\pr\!k}\ket{\hspace*{-1pt}n\!+\!1_{k}\hspace*{-1pt}}\!\ket{\hspace*{-1pt}m\!-\!1_{k\pr}\hspace*{-2pt}}\nonumber\\[1mm]
    &\ +\sqrt{n(m+1)}\,\Gpstar{k}\alphahmn{1}{kk\pr}\ket{\hspace*{-1pt}n\!-\!1_{k}\hspace*{-1pt}}\!\ket{\hspace*{-1pt}m\!+\!1_{k\pr}\hspace*{-2pt}}\nonumber\\[1mm]
    &\ -\sqrt{mn}\,\Gpstar{k}\betahmn{1}{kk\pr}\ket{\hspace*{-1pt}n\!-\!1_{k}\hspace*{-1pt}}\!\ket{\hspace*{-1pt}m\!-\!1_{k\pr}\hspace*{-2pt}}\nonumber\\[1mm]
    &\ -\sqrt{(n\!+\!1)(m\!+\!1)}\,\Gpstar{k\pr}\betahmnstar{1}{kk\pr}\ket{\hspace*{-1pt}n\!+\!1_{k}\hspace*{-1pt}}\!\ket{\hspace*{-1pt}m\!+\!1_{k\pr}\hspace*{-2pt}}\nonumber\\[1mm]
    &\qquad+\sum\limits_{p\neq k,k\pr}\Bigl(\sqrt{n}\,\Gpstar{k}\alphahmn{1}{kp}
    \ket{\hspace*{-1pt}n\!-\!1_{k}\hspace*{-1pt}}\!\ket{\hspace*{-1pt}m_{k\pr}\hspace*{-2pt}}\!\ket{\hspace*{-1pt}1_{p}\hspace*{-1pt}}\nonumber\\
    &\qquad\qquad+\sqrt{m}\,\Gpstar{k\pr}\alphahmn{1}{k\pr\!p}
    \ket{\hspace*{-1pt}n_{k}\hspace*{-1pt}}\!\ket{\hspace*{-1pt}m\!-\!1_{k\pr}\hspace*{-2pt}}\!\ket{\hspace*{-1pt}1_{p}\hspace*{-1pt}}\nonumber\\[1mm]
    &\qquad\qquad-\sqrt{n\!+\!1}\,\Gpstar{k}\betahmnstar{1}{pk}
    \ket{\hspace*{-1pt}n\!+\!1_{k}\hspace*{-1pt}}\!\ket{\hspace*{-1pt}m_{k\pr}\hspace*{-2pt}}\!\ket{\hspace*{-1pt}1_{p}\hspace*{-1pt}}\nonumber\\[1mm]
    &\qquad\qquad-\sqrt{m\!+\!1}\,\Gpstar{k\pr}\betahmnstar{1}{pk\pr}
    \ket{\hspace*{-1pt}n_{k}\hspace*{-1pt}}\!\ket{\hspace*{-1pt}m\!+\!1_{k\pr}\hspace*{-2pt}}\!\ket{\hspace*{-1pt}1_{p}\hspace*{-1pt}}\Bigr)\nonumber\\[1mm]
    &\qquad-\tfrac{1}{2}\!\!\sum\limits_{\substack{p,q\neq k,k\pr\\ p\neq q}}\Gpstar{q}\betahmnstar{1}{pq}
    \ket{\hspace*{-1pt}n_{k}\hspace*{-1pt}}\!\ket{\hspace*{-1pt}m_{k\pr}\hspace*{-2pt}}\ket{\hspace*{-1pt}1_{p}\hspace*{-1pt}}\!\ket{\hspace*{-1pt}1_{q}\hspace*{-1pt}}
    \nonumber\\
    &\qquad-\tfrac{1}{2}\!\!\sum\limits_{p\neq k,k\pr}\Gpstar{p}\betahmnstar{1}{pp}
    \ket{\hspace*{-1pt}n_{k}\hspace*{-1pt}}\!\ket{\hspace*{-1pt}m_{k\pr}\hspace*{-2pt}}\ket{\hspace*{-1pt}2_{p}\hspace*{-1pt}}
    \Bigr]+\mathcal{O}(\theta^{2}),
    \label{eq:appendix state transformation m plus n particles}
\end{align}
where $G_{m}^{*}\alphahmn{1}{mn}+G_{n}\alphahmnstar{1}{nm}=0$ and $G_{m}^{*}\betahmn{1}{mn}=G_{n}^{*}\betahmn{1}{nm}$ due to unitarity. For the initial state $\ket{\psi}=\ket{\hspace*{-1pt}n_{k}\hspace*{-1pt}}\!\ket{\hspace*{-1pt}m_{k\pr}\hspace*{-2pt}}$ we find
\begin{align}
    &\scpr{\!\tilde{\psi}^{\raisebox{-0.5pt}{\tiny{(1)}}}}{\tilde{\psi}^{\raisebox{-0.5pt}{\tiny{(1)}}}\!}   \,=\,
    \tfrac{1}{2}n(n+1)|\betahmn{1}{kk}|^{2}+\tfrac{1}{2}m(m+1)|\betahmn{1}{k\pr\!k\pr}|^{2}\nonumber\\[1mm]
    &+n^{2}|\alphahmn{1}{kk}|^{2}+m^{2}|\alphahmn{1}{k\pr\!k\pr}|^{2}+
    2mn\bigl(|\alphahmn{1}{kk\pr}|^{2}+|\betahmn{1}{kk\pr}|^{2}\bigr)
    \nonumber\\
    &+n\sum\limits_{p\neq k}\bigl(|\alphahmn{1}{pk}|^{2}+|\betahmn{1}{pk}|^{2}\bigr)+m\!\sum\limits_{p\neq k\pr}\bigl(|\alphahmn{1}{pk\pr}|^{2}
    +|\betahmn{1}{pk\pr}|^{2}\bigr)\nonumber\\
    &+\tfrac{1}{2}\sum\limits_{p,q}|\betahmn{1}{pq}|^{2}\,,
    \label{eq:appendix state transformation m plus n particles psi 1 psi 1}
\end{align}
\vspace*{-4mm}
\begin{align}
    \mbox{and}\qquad|\!\scpr{\!\tilde{\psi}^{\raisebox{-0.5pt}{\tiny{(0)}}}}{\tilde{\psi}^{\raisebox{-0.5pt}{\tiny{(1)}}}\!}\!|^{2}  &=\,
    n^{2}|\alphahmn{1}{kk}|^{2}+m^{2}|\alphahmn{1}{k\pr\!k\pr}|^{2}\,.
    \label{eq:appendix state transformation m plus n particles psi 0 psi 1}
\end{align}
We hence obtain the QFI for the state $\ket{\hspace*{-1pt}n_{k}\hspace*{-1pt}}\!\ket{\hspace*{-1pt}m_{k\pr}\hspace*{-2pt}}$ to be
\begin{align}
    &\mathcal{I}(\ket{\hspace*{-1pt}n_{k}\hspace*{-1pt}}\!\ket{\hspace*{-1pt}m_{k\pr}\hspace*{-2pt}})   \,=\,
    2n(n+1)|\betahmn{1}{kk}|^{2}+2m(m+1)|\betahmn{1}{k\pr\!k\pr}|^{2}\nonumber\\[1mm]
    &+8mn\bigl(|\alphahmn{1}{kk\pr}|^{2}+|\betahmn{1}{kk\pr}|^{2}\bigr)
    +4n\sum\limits_{p\neq k}\bigl(|\alphahmn{1}{pk}|^{2}+|\betahmn{1}{pk}|^{2}\bigr)
    \nonumber\\
    &+4m\!\!\sum\limits_{p\neq k\pr}\bigl(|\alphahmn{1}{pk\pr}|^{2}
    +|\betahmn{1}{pk\pr}|^{2}\bigr)+2\!\!\sum\limits_{p,q}|\betahmn{1}{pq}|^{2}+\mathcal{O}(\theta)\,.
    \label{eq:appendix QFI m plus n particles}
\end{align}
\ \\
\vspace*{-4mm}

\subsection{Optimal scaling}
\vspace*{-2mm}

Finally, we show that, within the perturbative approach we pursue here, no better scaling of the QFI is possible. To see this, note that Heisenberg scaling arises from the linear order correction terms in Eq.~(\ref{eq:appendix state transformation m plus n particles}) that are proportional to square roots of at most quadratic combinations of~$m$ and~$n$. Each of these terms corresponds to either~a shift of excitations between the modes~$k$ and~$k\pr$, indicated by coefficients~$\alphahmn{1}{kk\pr}$, or the creation (annihilation) of~a pair of particles in these modes, corresponding to the coefficients~$\betahmn{1}{kk}$,~$\betahmn{1}{kk\pr}$, and~$\betahmn{1}{k\pr\!k\pr}$. In other words, terms in~$\ket{\tilde{\psi}^{\raisebox{-0.5pt}{\tiny{(1)}}}\!}$ that scale as~$n$ are the result of changing the occupation number of two highly excited modes by one excitation each or the occupation of one such mode twice. To linear order in~$\theta$, no more than these two excitations may be changed. No initial state may hence have~a linear correction~$\ket{\tilde{\psi}^{\raisebox{-0.5pt}{\tiny{(1)}}}\!}$ that grows faster than linearly in~$n$.

This still leaves the option to consider superpositions of states with different occupation numbers, such that the average number of excitations remains fixed, for instance, the state $\bigl(\ket{\hspace*{-1pt}n_{k}\hspace*{-1pt}}\!\ket{\hspace*{-1pt}n_{k\pr}\hspace*{-2pt}}
+\ket{\hspace*{-1pt}n_{k}\hspace*{-1pt}}\!\ket{\hspace*{-1pt}n\!-\!2_{k\pr}\hspace*{-2pt}}+\ket{\hspace*{-1pt}n_{k}\hspace*{-1pt}}\!\ket{\hspace*{-1pt}n\!+\!2_{k\pr}\hspace*{-2pt}}\bigr)/\sqrt{3}$. The cross terms appearing in $\scpr{\!\tilde{\psi}^{\raisebox{-0.5pt}{\tiny{(1)}}}}{\tilde{\psi}^{\raisebox{-0.5pt}{\tiny{(1)}}}\!}$ may increase (or decrease) the overall QFI. However, to linear order in~$\theta$ the corrections to every fixed-excitation state in the superposition are nonorthogonal only to the linear corrections to states that differ by no more than two excitations. For example, the linear corrections to both~$\ket{\hspace*{-1pt}n_{k}\hspace*{-1pt}}\!\ket{\hspace*{-1pt}n_{k\pr}\hspace*{-2pt}}$ and $\ket{\hspace*{-1pt}n_{k}\hspace*{-1pt}}\!\ket{\hspace*{-1pt}n\!-\!2_{k\pr}\hspace*{-2pt}}$ contain~a term proportional to \ket{\hspace*{-1pt}n\!-\!1_{k}\hspace*{-1pt}}\!\ket{\hspace*{-1pt}n\!-\!1_{k\pr}\hspace*{-2pt}}, whereas the linear correction of \ket{\hspace*{-1pt}n_{k}\hspace*{-1pt}}\!\ket{\hspace*{-1pt}n\!+\!2_{k\pr}\hspace*{-2pt}} cannot contain such~a term, as can easily be verified from Eq.~(\ref{eq:appendix state transformation m plus n particles}). The possible gain in the number of such cross terms that is obtained from adding more terms to the superposition is therefore at most linear in the number of terms, which is compensated by the requirement of normalization. Consequently, the leading-order correction (in $\theta$) to the QFI is at most quadratic in the average occupation number, as we have claimed.

The higher-order corrections may feature higher powers of~$n$. However, the perturbative nature of the calculations means that the results presented here are only valid as long as the higher-order corrections are negligibly small with respect to the leading order. Therefore, the perturbative approach is not uniformly valid in the average occupation number. In~a regime where~$n$ is large enough such that $\theta^{2}\mathcal{I}(\ket{\psi})/4\approx1$ the perturbative calculation no longer yields~a reliable result. We can therefore conclude that, within the perturbative regime, Heisenberg scaling is indeed optimal.

\end{document}